\documentclass[prl,amsmath,amssymb,twocolumn,showpacs,floatfix]{revtex4}
\usepackage{graphicx}
\usepackage{dcolumn}
\usepackage{bm}

\begin{document}

\noindent{\bf Comment on ``Magnetic Order in the Pseudogap Phase of High-$T_c$ Superconductors"}

In a recent Letter, Fauqu\'{e} {\it et al.} \cite{Fauque:06} reported the detection of
a novel magnetic order in YBa$_2$Cu$_3$O$_y$ (YBCO) near the pseudogap transition temperature 
$T^*$ by polarized neutron diffraction. They remark that this
magnetic order has not been clearly identified by local probes and, in particular, by
muon spin resonance ($\mu$SR). The purpose of this Comment is to point out that this
is untrue, and together with the more recent detection of broken time reversal
symmetry by polar Kerr effect measurements \cite{Xia:08}, there are now three very different 
kinds of experiments that provide evidence for this magnetic transition.    

The onset of weak static magnetism near $T^*$ was in fact first detected in 
YBCO by zero-field (ZF) $\mu$SR \cite{Sonier:01}. 
A subsequent study \cite{Sonier:02} revealed 
correlations with charge inhomogeneities and structural changes that
may play a role in stabilizing the magnetism.
It has already been mentioned in Ref.~\cite{Xia:08} that the polar Kerr effect and 
ZF-$\mu$SR measurements show similar magnetic onset temperatures
for $y \! = \! 6.67$ and $y \! = \! 6.95$ single crystals. However, since
the hole doping $p$ in the CuO$_2$ layers of YBCO not only depends on 
the oxygen content $y$, but also on the
degree of oxygen ordering in the CuO chains, a proper comparison of the data
from different techniques should be between samples with the same value of $p$.
In Ref.~\cite{Sonier:01}, $p$ was calculated
from the superconducting transition temperature $T_c$ using an empircal parabolic equation 
deduced from data on La$_{2-x}$Sr$_x$CuO$_4$ \cite{Presland:91}. Recently, a more accurate 
relationship between $p$ and $T_c$ has been established for YBCO that 
properly accounts for the suppression of $T_c$ near $p \! = \! 1/8$ \cite{Liang:06}.

Figure~1 shows a comparison of the magnetic onset temperatures identified by
the three different techniques, where the appropriate empirical relationship from 
Ref.~\cite{Liang:06} has been used to determine $p$ for all data sets.
It is reasonable to conclude from this plot that there is a single magnetic 
transition---{\it i.e.} the magnetism detected earlier
by ZF-$\mu$SR originates from the same source as the magnetic order detected by
polarized neutron diffraction. While the neutron data is consistently above the
ZF-$\mu$SR and polar Kerr effect data, Fig.~1 does not account for the transition 
widths $\Delta T_c$ ($\Delta p$), which are typically broader in the larger neutron
samples.                

\begin{figure}
\centering
\includegraphics[width=9.0cm]{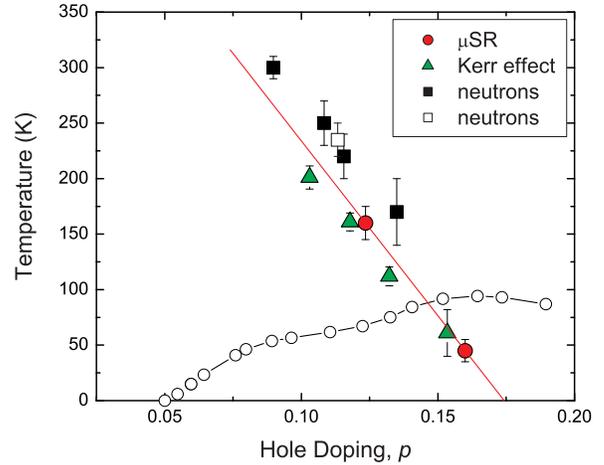}
\caption{(color online) Hole doping dependence of the onset temperature for
anomalous magnetic order in pure YBCO from Ref.~\cite{Fauque:06} (solid squares),
Ref.~\cite{Xia:08} (green triangles), Ref.~\cite{Sonier:01} (red circles),
and Ref~\cite{Mook:08} (open square). The solid circles are $T_c$ from Ref.~\cite{Liang:06}.
The red line is a fit to the $\mu$SR data, which
vanishes at $p \! = \! 0.174$.}
\label{fig1}
\end{figure}

The size of the local fields $B_{\rm loc}$ detected by ZF-$\mu$SR depends 
on the type of magnetic order and the muon stopping sites. 
For $y \! = \! 6.95$, nearly all of the muons
stop at a distance of $\sim \! 1$~\AA~ from a chain oxygen \cite{Pinkpank:99} at
the position (0.15(1), 0.44(1), 0.071(1)) \cite{Weber:90}, whereas $\sim \! 1/3$
of the muons stop near an apical oxygen for $y \! = \! 6.67$. Assuming
either the orbital current pattern or spin model displayed in Fig.~3 of 
Ref.~\cite{Fauque:06}, a simple dipolar-field calculation yields 
$B_{\rm loc} \! \leq \! 130$~G/$\mu_B$ at the chain oxygen site. 
Fauqu\'{e} {\it et al.} report an ordered magnetic moment of 0.05 to 0.1~$\mu_B$ 
that decreases with increasing $p$ for $p \! \sim \! 0.09$ to 0.135. 
Consequently, for $y \! = \! 6.95$ ($p \! = \! 0.16$) a field of 
less than 6.5~G should be detected by ZF-$\mu$SR. This is 
not inconsistent with the measured characteristic field of $\sim \! 0.3$~G \cite{Sonier:01},
especially since: (i) the proposed orbital and spin models do not account for  
the angle of $45 \! \pm \! 20^\circ$ the ordered moment makes with the $c$ axis 
\cite{Fauque:06,Mook:08}, and (ii) the relaxation rate of the ZF-$\mu$SR signal
is sufficiently weak that one cannot say (even from the other techniques) 
whether the magnetic order occurs in the full sample volume. 
If it does not, then the ZF-$\mu$SR experiments
give a lower limit for the static local field in the magnetic regions.\\ 
       
\noindent Jeff E. Sonier \\
Department of Physics \\
Simon Fraser University \\
Burnaby, British Columbia \\
Canada V5A 1S6 \\


\begin{thebibliography}{12} 

\bibitem{Fauque:06} B.~Fauqu\'{e} {\it et al.}, Phys.~Rev.~Lett. {\bf 96}, 197001 (2006).

\bibitem{Xia:08} J.~Xia {\it et al.}, Phys.~Rev.~Lett. {\bf 100}, 127002 (2008).

\bibitem{Sonier:01} J.E.~Sonier {\it et al.}, Science {\bf 292}, 1692 (2001).

\bibitem{Sonier:02} J.E.~Sonier {\it et al.}, Phys.~Rev.~B. {\bf 66}, 134501 (2002).

\bibitem{Presland:91} M.E.~Presland {\it et al.}, Physica~C {\bf 176}, 95 (1991).

\bibitem{Liang:06} R.~Liang, D.A.~Bonn, and W.N.~Hardy, Phys.~Rev.~B. {\bf 73}, 180505(R) (2006).

\bibitem{Mook:08} H.~Mook {\it et al.}, arXiv:0802.3620.

\bibitem{Pinkpank:99} M.~Pinkpank {\it et al.}, Physica~C {\bf 317-318}, 299 (1999).

\bibitem{Weber:90} M.~Weber {\it et al.}, Hyperfine Interactions {\bf 63}, 207 (1990).  

\end{thebibliography}
\end{document}